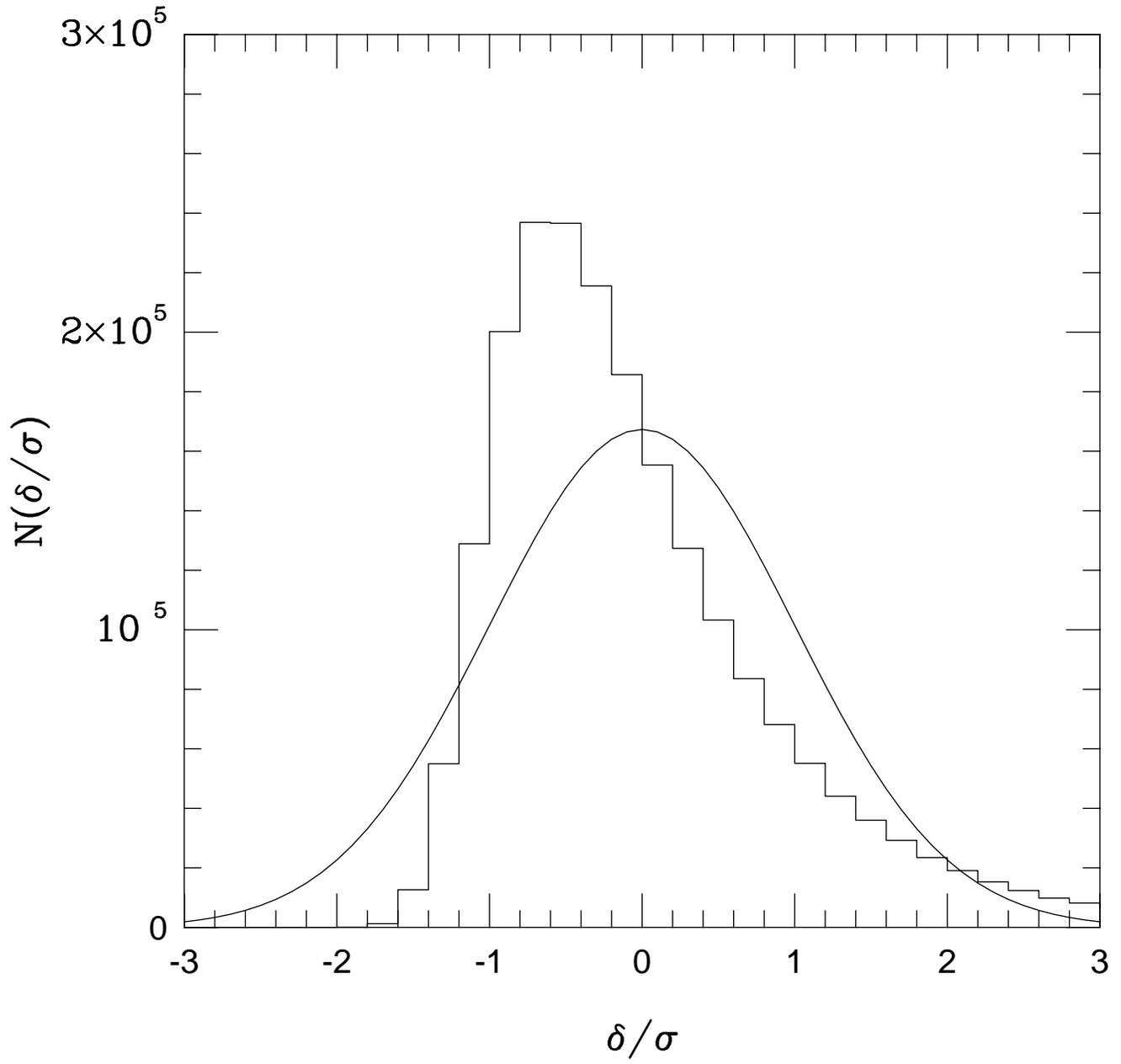

Figure 1 (a)

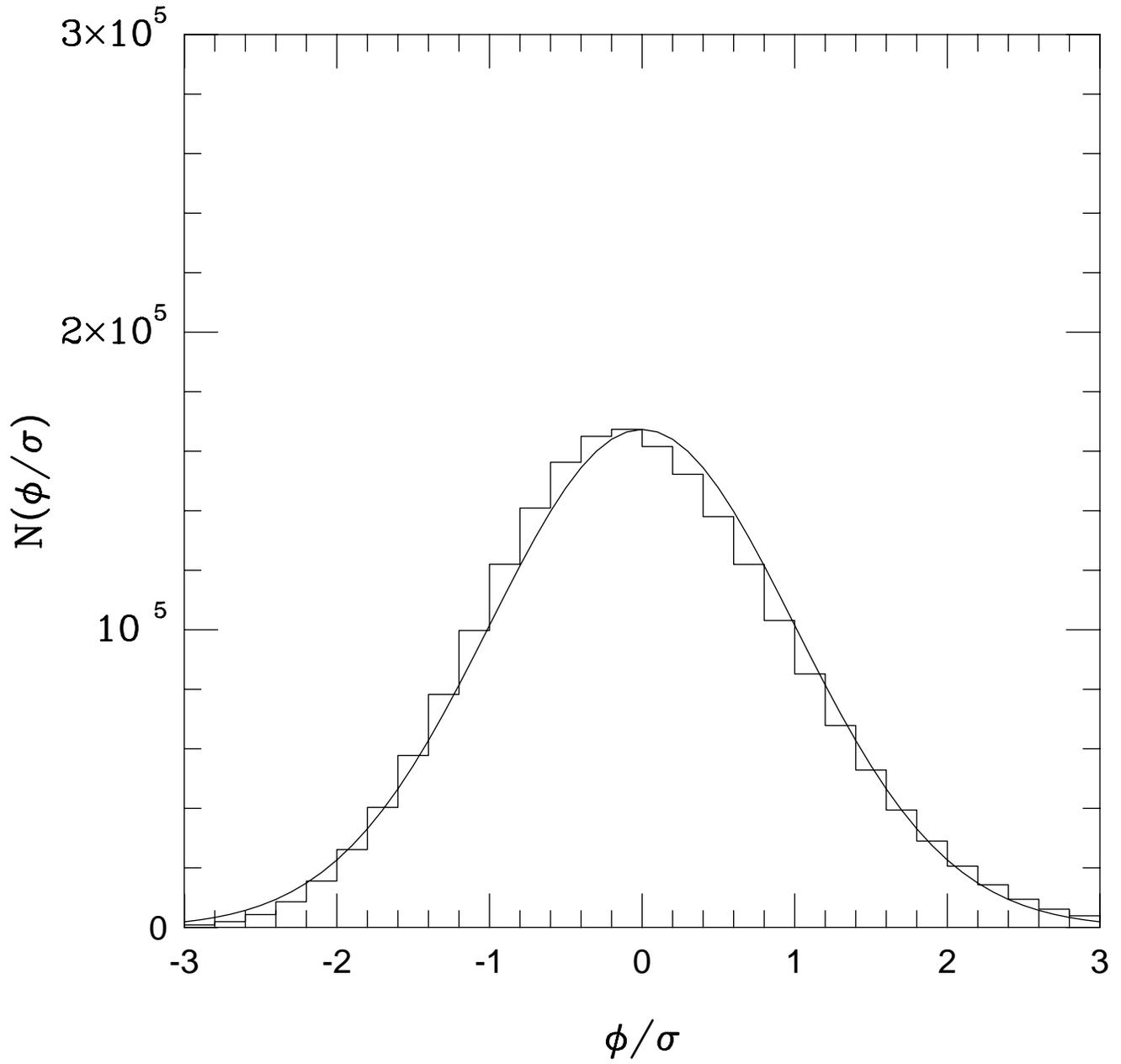

Figure 1 (b)

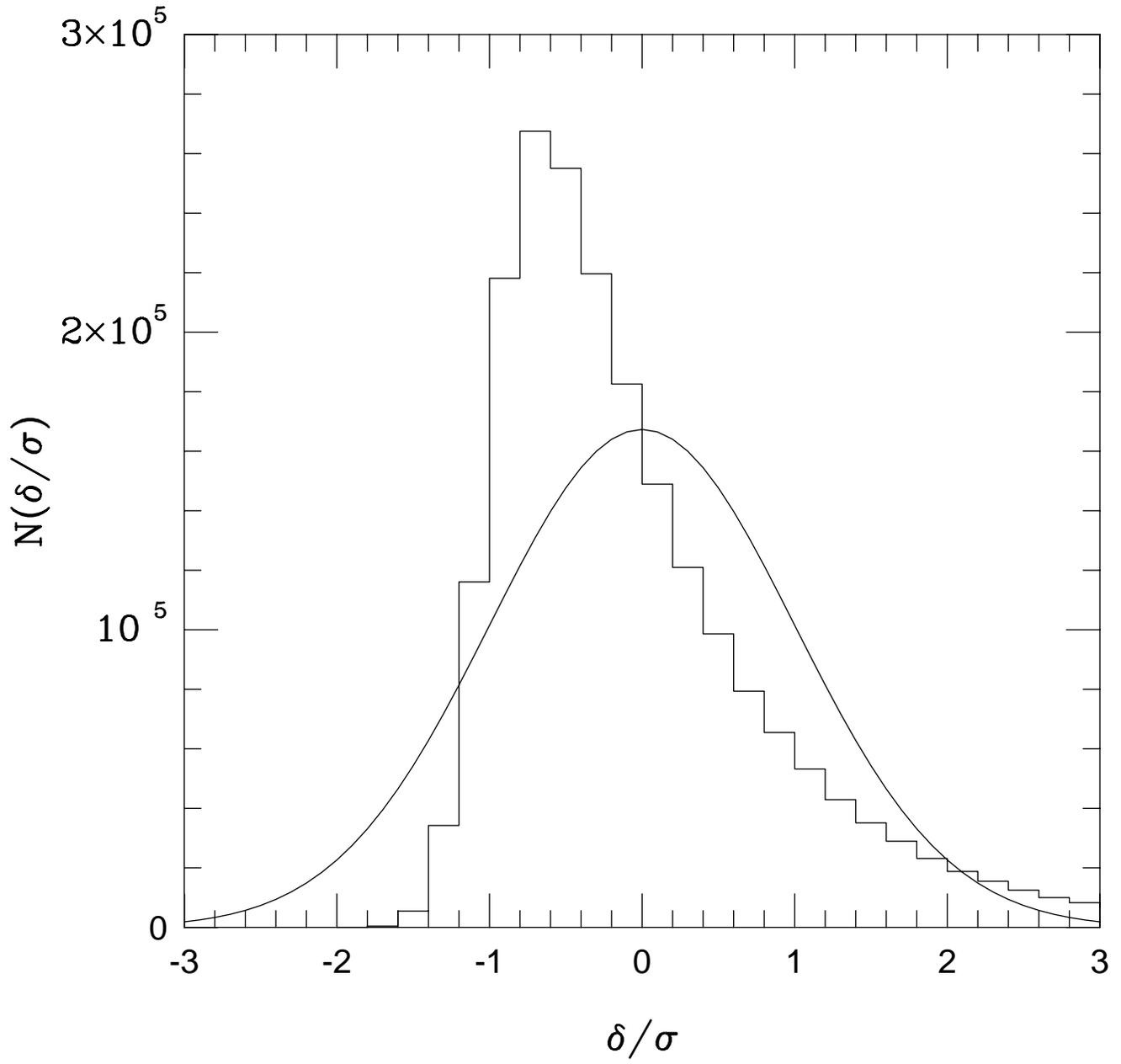

Figure 2 (a)

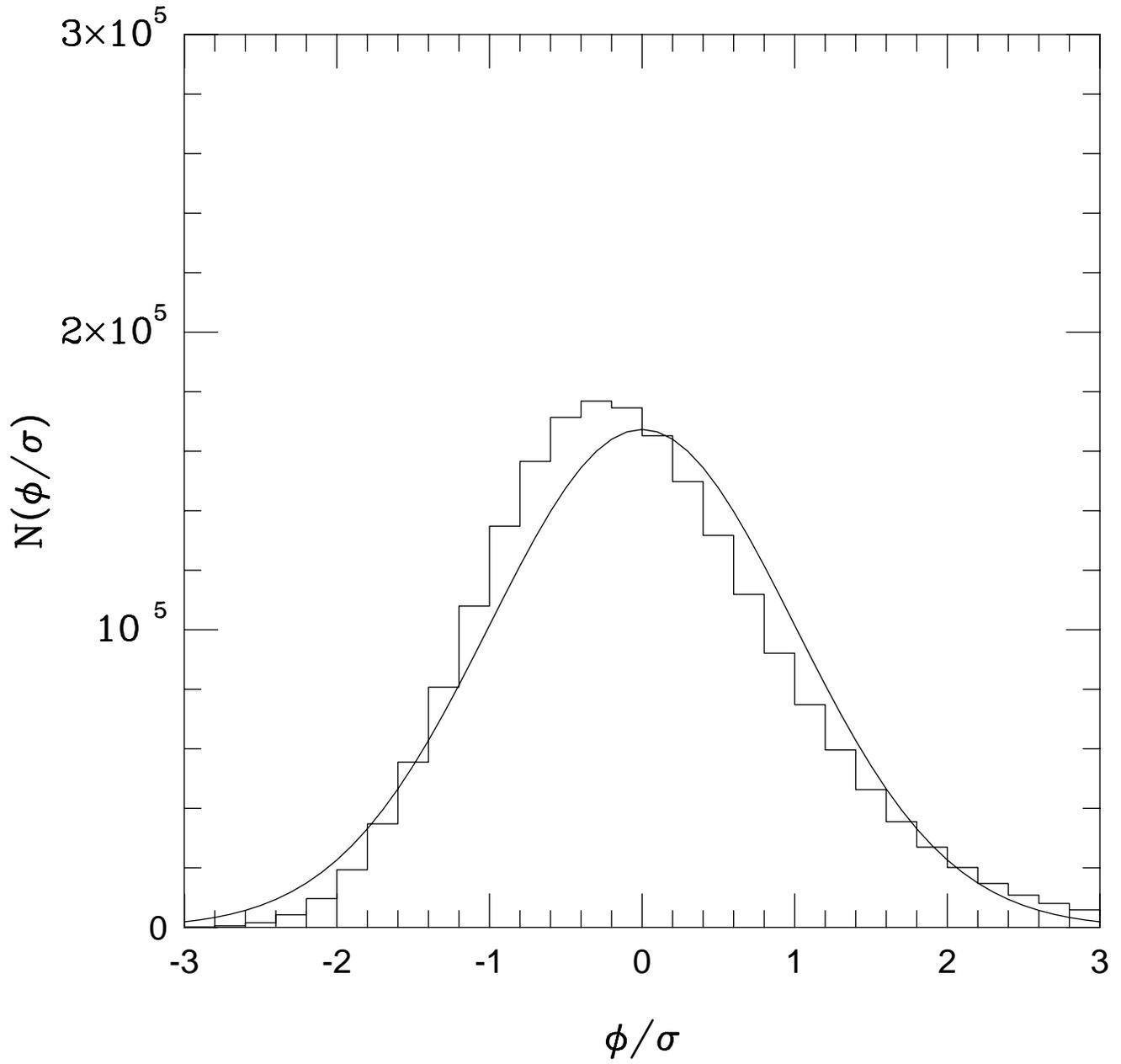

Figure 2 (b)

# WHEN CAN NON-GAUSSIAN DENSITY FIELDS PRODUCE A GAUSSIAN SACHS-WOLFE EFFECT?

Robert J. Scherrer

Department of Physics, The Ohio State University

Columbus, OH 43210

Robert K. Schaefer

Bartol Research Institute, University of Delaware

Newark, DE 19716

## Abstract

The Sachs-Wolfe temperature fluctuations produced by primordial density perturbations are proportional to the potential field $\phi$, which is a weighted integral over the density field $\delta$. Because of the central limit theorem, $\phi$ can be approximately Gaussian even when $\delta$ is non-Gaussian. Using the Wold representation for non-Gaussian density fields, $\delta(\mathbf{r}) = \int f(|\mathbf{r} - \mathbf{r}'|)\Delta(\mathbf{r}')d^3\mathbf{r}'$, we find conditions on $\Delta$ and $f$ for which $\phi$ must have a Gaussian one-point distribution, while $\delta$ can be non-Gaussian. Sufficient (but not necessary) conditions are that the density field have a power spectrum (which determines $f$) of $P(k) \propto k^n$, with $-2 < n \le +1$, and that $\Delta(\mathbf{r})$ be non-Gaussian with no long-range correlations. Thus, there is an infinite set of non-Gaussian density fields which produce a nearly Gaussian one-point distribution for the Sachs-Wolfe effect.

The COBE observations of Sachs-Wolfe fluctuations in the cosmic microwave background place severe constraints on possible models for large-scale structure (Smoot et al. 1992; Wright et al. 1992). One of the unresolved questions regarding the primordial density fluctuations which gave rise to large-scale structure is whether these density fluctuations were Gaussian, i.e., whether the N-point distribution of primordial density fluctuations is a multivariate Gaussian distribution. Gaussian models have the virtue of simplicity, and a great deal is known about their properties (Bardeen, Bond, Kaiser, & Szalay 1986); on the other hand, a large number of non-Gaussian models have also been proposed. Consequently, many recent papers have addressed the issue of looking for a non-Gaussian signal in the COBE data (Kung 1993; Srednicki 1993; Luo & Schramm 1993; Luo 1994a,b; Hinshaw et al. 1994; Smoot et al. 1994), as well as predicting the microwave fluctuations to be expected in non-Gaussian models (Bennett & Rhie 1993; Coulson, et al. 1994; Pen & Spergel 1994; Moessner, Perivolaropoulos, & Brandenberger 1994).

However, Sachs-Wolfe observations cannot provide a decisive test of whether or not the initial density field is Gaussian; in this paper, we derive conditions under which a non-Gaussian density field can produce a Gaussian one-point temperature distribution. We mention first the well-known possibility that the density field could be non-Gaussian on small scales but Gaussian on the length scales probed by COBE. Such density fields are produced, for example, by a Poisson distribution of objects; the density field is non-Gaussian when smoothed on scales smaller than the mean separation of objects, but Gaussian when smoothed on scales larger than this mean separation (Scherrer & Bertschinger 1991).

We can, however, make a stronger assertion: the potential field can have a Gaussian one-point distribution when smoothed on a given length scale even when the density field is non-Gaussian on this same scale, even in the limit where the smoothing length goes to 0. The Sachs-Wolfe fluctuations probed by COBE are directly proportional to the potential field $\phi(\mathbf{r})$ on the surface of last scattering. This potential field is related to the field of density fluctuations $\delta(\mathbf{r})$ via

$$\nabla^2 \phi(\mathbf{r}) = 4\pi G a^2 \bar{\rho} \delta(\mathbf{r}), \tag{1}$$

where $a$ is the cosmological scale factor and $\bar{\rho}$ is the mean density. [Although the Sachs-Wolfe effect is inherently relativistic, our use of the Newtonian expressions for the potential will give the correct answer for the problem of interest here. This is justified in the Appendix]. In integral form, we can write this relationship as:

$$\phi(\mathbf{r}) = -Ga^2\bar{\rho}\int \frac{\delta(\mathbf{r}')d^3\mathbf{r}'}{|\mathbf{r}-\mathbf{r}'|}. \quad (2)$$

Thus, the potential field is a weighted integral over the density field. The central limit theorem indicates that this integral can be driven to a Gaussian distribution even when the distribution of $\delta$ is non-Gaussian. A similar phenomenon has been noted in the case of the velocity field, which can also be expressed as an integral over the density field; the distribution of velocities can be Gaussian even when the underlying density field is non-Gaussian (Scherrer 1992).

The first step in our argument is to express the density field in terms of the Wold representation (Scargle 1981; Peebles 1983). Any density field can be represented as the convolution of a spherically symmetric function $f$ with a white-noise function $\Delta$:

$$\delta(\mathbf{r}) = \int f(|\mathbf{r}-\mathbf{r}'|)\Delta(\mathbf{r}')d^3\mathbf{r}'. \quad (3)$$

The function $\Delta$ is a white noise function in the sense that the two-point correlation function is zero: $\langle\Delta(\mathbf{r}_1)\Delta(\mathbf{r}_2)\rangle = 0$ for $\mathbf{r}_1 \neq \mathbf{r}_2$. However, $\Delta$ can (and in general will) have non-zero higher-order correlations. The proof that any density field can be represented in the form of equation (3) is trivial. The Fourier decomposition of $\delta(\mathbf{r})$ gives us a set of Fourier components $\hat{\delta}(\mathbf{k})$. If we write each $\hat{\delta}(\mathbf{k})$ as $\hat{\delta}(\mathbf{k}) = R(\mathbf{k})e^{i\theta(\mathbf{k})}$, where $R$ and $\theta$ are real, then the Fourier transform of $R$ gives $f$ (which must be spherically symmetric because $R$ is real) and the Fourier transform of $e^{i\theta}$ gives $\Delta$ (which has a white noise power spectrum, since $|e^{i\theta(\mathbf{k})}|^2 = 1$, but which may have non-zero higher-order correlations). In a sense, the Wold decomposition is the most natural representation for non-Gaussian models. The phase correlations in the model are contained entirely in the function $\Delta$, while the power spectrum is given entirely by the function $f$. For example, the phase correlation statistic of Scherrer, Melott & Shandarin (1991) involves calculating the properties of the field $\Delta(\mathbf{r})$. A number of authors have used the Wold representation to generate non-Gaussian fields (Peebles 1983; Lucchin & Matarrese 1988; Moscardini et al. 1991).

We will assume that the density field on the scales probed by COBE has a power spectrum $P(k) \propto k^n$. The Fourier components of the density and potential field are given by

$$\hat{\delta}(k) \propto k^{n/2}\hat{\Delta}(k), \tag{4}$$

and

$$\hat{\phi}(k) \propto k^{n/2-2}\hat{\Delta}(k), \tag{5}$$

where $\hat{\Delta}(k)$ is the Fourier transform of $\Delta(\mathbf{r})$ which appears in equation (3). To derive an expression for $f(|\mathbf{r} - \mathbf{r}'|)$ in the Wold representation of the density field, we Fourier transform $k^{n/2}$ to obtain

$$\delta(\mathbf{r}) \propto \int \Delta(\mathbf{r}')|\mathbf{r} - \mathbf{r}'|^{-n/2-3} d^3\mathbf{r}'. \tag{6}$$

[This Fourier transform of a power-law in $k$ is discussed by Peebles (1980). Strictly speaking, convergence of the Fourier transform here requires $n/2 > -3$ and smoothing at small scales; equation (6) is then the correct expression for $|\mathbf{r} - \mathbf{r}'|$ greater than the smoothing scale. Also, equation (6) fails for the white noise case $n = 0$; in that case we simply have $\delta(\mathbf{r}) = \Delta(\mathbf{r})$.] Equation (1) or (2) tells us that $\phi$ can be written as

$$\phi(\mathbf{r}) \propto \int \Delta(\mathbf{r}')|\mathbf{r} - \mathbf{r}'|^{-n/2-1} d^3\mathbf{r}'. \tag{7}$$

[Again, this expression is valid only for $n/2 > -1$, which is satisfied by the COBE results (Wright, et al. 1994; Gorski et al. 1994). The case $n = +4$ corresponds to a white-noise potential distribution, for which $\phi(\mathbf{r}) \propto \Delta(\mathbf{r})$.]

We will assume that the density field and potential field are smoothed with some window function of radius $R$, where $R$ can be arbitrarily small. Note that since smoothing is a convolution, we can simply smooth $\Delta(\mathbf{r})$ with this window function to give the correctly-smoothed density and potential fields. Now we have to make some assumption regarding the nature of $\Delta$. For now, we will assume that $\Delta$ has a non-Gaussian one-point distribution $p(\Delta)$, when smoothed over the length scale $R$, but that all higher-order correlations for $r > R$ vanish. In a sense, this is a minimally non-Gaussian model; such models (in which $\Delta$ was used as the density field) were investigated numerically by Messina et al. (1990). This also describes a special class of the seed models investigated by Scherrer & Bertschinger (1991) and Scherrer (1992), namely, seed models in which all of the seeds have identical mass and are uncorrelated. In this case $p(\Delta)$ is just a discrete probability distribution and $f$ is the accretion pattern around each seed.

We can now apply the central limit theorem for the sum of weighted independent random variables (Feller 1971, p. 264). Consider a random variable $X$ with a known distribution, having zero mean and unit variance. Now construct a set of independent random variables $X_k$ given by $X_k = \sigma_k X$. If the random variable $Z_N$ is the sum of the first $N$ $X_k$'s:

$$Z_N = \sum_{k=1}^{N} X_k, \qquad (8)$$

and $s_N$ is the variance of $Z_N$:

$$s_N^2 = \sum_{k=1}^{N} \sigma_k^2, \qquad (9)$$

then $Z_N/s_N$ has a Gaussian distribution in the limit $N \to \infty$ if both

$$s_N^2 \to \infty, \qquad (10)$$

and

$$\sigma_N/s_N \to 0. \qquad (11)$$

By smoothing on a length scale $R$, we have effectively discretized the potential and density fields. Think of the $\Delta$ field as divided into discrete regions of volume $\sim R^3$. Then equations (6) and (7) can be written as sums, where we translate to the origin:

$$\delta(0) \propto \sum_{k=1}^{\infty} \Delta(\mathbf{r}_k) |\mathbf{r}_k|^{-n/2-3} R^3, \qquad (12)$$

and

$$\phi(0) \propto \sum_{k=1}^{\infty} \Delta(\mathbf{r}_k) |\mathbf{r}_k|^{-n/2-1} R^3. \qquad (13)$$

We choose an order of summation such that we first sum over all regions at a given radius $|\mathbf{r}|$, then move on to the next larger radius. This gives $|\mathbf{r}_k| \propto k^{1/3}$, so $\sigma$ for the $j^{th}$ term in each sum is

$$\sigma_j \propto \sigma_\Delta j^{-n/6-1}, \qquad (14)$$

for the density field, and

$$\sigma_j \propto \sigma_\Delta j^{-n/6-1/3}, \tag{15}$$

for the potential field, where $\sigma_\Delta^2$ is the variance of the $\Delta$ field. Then the quantity $s_N^2$ is just

$$s_N^2 \propto \sigma_\Delta^2 \sum_{k=1}^{N} k^{-n/3-2}, \tag{16}$$

for the density field, and

$$s_N^2 \propto \sigma_\Delta^2 \sum_{k=1}^{N} k^{-n/3-2/3} \tag{17}$$

for the potential field. Consider first the case of a Zel'dovich power spectrum ($n = +1$). For the density field, we note that $s_N^2 \propto \sum_{k=1}^{N} k^{-7/3}$, which converges to a finite value as $N \to \infty$. Thus, the central limit theorem will not apply to the density field in this case. For the potential field, we have $s_N^2 \propto \sum_{k=1}^{N} k^{-1}$, so that $s_N \to \infty$ as $N \to \infty$. Furthermore, $\sigma_N \propto N^{-1/2}$ so $\sigma_N/s_N \to 0$ as $N \to \infty$. Thus, the central limit theorem is satisfied, and the potential field will have a Gaussian distribution.

We can ask the more general question: what are the limits on $n$ for which $\phi$ must be Gaussian (independent of the one-point distribution of $\Delta$), while $\delta$ can be non-Gaussian? We note that the sum $\sum_k k^s$ converges for $s < -1$ and diverges otherwise. Then equation (17) tells us that for the potential field, $s_N \to \infty$ for $N \to \infty$ as long as $n \leq +1$, while for $n < +1$, $\sigma_N/s_N \sim 1/\sqrt{N}$ for large $N$, so equation (11) is also satisfied. Conversely, if we wish to violate equation (10) for the density field, then equation (16) tells us that $n > -3$. These results make intuitive sense. For $-3 < n < +1$, the integral over the power spectrum which gives $\langle \delta^2 \rangle$ is dominated by large $k$ (small length scales), while the corresponding integral for $\langle \phi^2 \rangle$ is dominated by small $k$ (large length scales), so it is reasonable that the central limit theorem works for $\phi$ but not for $\delta$ in these cases. Since equation (7) is valid only for $n > -2$, we can conclude that for a power spectrum of the form $P(k) \propto k^n$, with $-2 < n \leq +1$, there is an infinite class of models for which the density field can be highly non-Gaussian, while the potential field has a nearly Gaussian one-point distribution. This range includes the Zel'dovich power spectrum and is consistent with part of the allowed range given by COBE (Wright et al. 1994; Gorski et al. 1994). Note that $\sigma_\Delta$ must be finite for our argument to be valid, although no physically-realistic models have been proposed which produce an infinite variance for the density and potential fields.

To illustrate this result, we have generated two sets of density and potential fields with different power spectra, but the same non-Gaussian distribution for $\Delta(\mathbf{r})$. We begin by setting down the field $\Delta(\mathbf{r})$ on a $128^3$ lattice with a gamma function one-point distribution $p(\Delta) \propto \Delta e^{-\Delta}$, and no correlations between the values of $\Delta$ on different lattice sites. We Fourier transform $\Delta(\mathbf{r})$ to obtain $\hat{\Delta}(\mathbf{k})$, and then we take $\hat{\delta}(\mathbf{k}) = k^{1/2}\hat{\Delta}(\mathbf{k})$ and $\hat{\phi}(\mathbf{k}) = k^{-3/2}\hat{\Delta}(\mathbf{k})$ for Fig. 1, and $\hat{\delta}(\mathbf{k}) = k^1 \hat{\Delta}(\mathbf{k})$ and $\hat{\phi}(\mathbf{k}) = k^{-1}\hat{\Delta}(\mathbf{k})$ for Fig. 2. (Multiplicative constants are irrelevant for our purposes). We then inverse Fourier transform $\hat{\delta}(\mathbf{k})$ and $\hat{\phi}(\mathbf{k})$ to obtain the density field $\delta(\mathbf{r})$ and the corresponding potential field $\phi(\mathbf{r})$. To eliminate lattice effects, we smooth both of these fields with a Gaussian window function $W(r) = e^{-r^2/2R^2}$, with $R = 3$ cell lengths.

Our figures give the one-point distribution of the density and potential fields for $P(k) \propto k^n$ with $n = 1$ (Fig. 1) and $n = 2$ (Fig. 2). The distributions of both density fields (Figs. 1a and 2a) closely resemble the original gamma function. This is not suprising, since in both cases the form for $f(|\mathbf{r} - \mathbf{r}'|)$ in the Wold representation (equation 3) is sharply peaked near zero. The corresponding distributions for the potential fields are shown in Figs. 1b and 2b. In the Zel'dovich case (Fig. 1b) the distribution of the potential field retains a slight imprint of the original gamma distribution, but it is nearly indistinguishable from a Gaussian, confirming our argument that the potential field will be driven to a Gaussian in this case. Fig. 2b does not satisfy our conditions for the central limit theorem to apply, and the distribution of the potential in this case is obviously non-Gaussian.

Although we have derived our results for the case where $\Delta$ is uncorrelated, our argument generalizes to the case where $\Delta$ has short-range correlations, but is uncorrelated over distances greater than some finite length (Ibragimov & Linnik 1971). Suppose that $\Delta$ has non-zero correlations for separations less than some length $r_0$, but is uncorrelated over distances greater than $r_0$. Then we can group terms in equation (13) into regions with some size $r \gg r_0$, and let $\varphi_j$ be the contribution to the sum in equation (13) from the $j^{th}$ region: $\phi = \sum_j \varphi_j$. There will be correlations between the $\varphi_j$'s for adjacent regions, but, since $r \gg r_0$, there will be no correlations between regions which are not adjacent. Now we break the sum in equation (13) into two pieces, $\phi = \phi_1 + \phi_2$, dividing our regions in checkerboard fashion, so that no two regions in the sum for $\phi_1$ are adjacent, and no two regions in the sum for $\phi_2$ are adjacent. The central limit theorem applies to both $\phi_1$ and $\phi_2$; both must have a Gaussian distribution. Now we fix the size of the regions for $\phi_2$, but increase the size of the regions for $\phi_1$ to arbitrarily large size; the result is that $\phi \approx \phi_1$, so $\phi$ has a Gaussian distribution.

Note that this argument does *not* require smoothing: it applies even to the case where the smoothing length is much smaller than the maximum length over which $\Delta$ is correlated. [In the opposite case, the argument is trivial.] This is a special case of a more general result given by Ibragimov & Linnik (1971), who show that the central limit theorem applies if the correlations go to zero sufficiently rapidly with increasing separation. Thus, it is likely that our results can be generalized further, to sufficiently weakly correlated $\Delta$, although what "sufficiently weakly" means in this context is unclear.

Our results have several limitations. As we have noted, they do not apply to cases where $\Delta$ has long-range correlations or infinite variance. Of course, our results cannot be completely general, since it is obviously possible to choose the potential field to be non-Gaussian by construction. Furthermore, our results apply only to models in which the dominant cosmic microwave fluctuations are proportional to the potential at the surface of last scattering, not to models in which secondary effects are important. Finally, we have shown only that the one-point distribution function for the potential field is Gaussian, although it may be possible to generalize our results to the full N-point distribution function. [Non-Gaussian models with a Gaussian one-point distribution are certainly possible, but somewhat contrived (Feller 1971; Scherrer, Melott, & Shandarin 1991).] Within these limitations, however, our results do indicate that there is an infinite set of models for which the density field has a non-Gaussian distribution, but the potential field has a Gaussian one-point distribution, even when smoothed on arbitrarily small scales. Thus, even if the cosmic microwave background turns out to have a Gaussian one-point distribution on a given scale, this observation by itself cannot prove that the underlying density field is Gaussian on that scale.

R.J.S. was supported by NASA (NAGW-2589) and by the Department of Energy

(DE-AC02-76ER01545). R.K.S. was supported by NASA (NRA-91-OSSA-11).

## APPENDIX

In calculating the temperature fluctuations due to the Sachs-Wolfe effect, we must evaluate the gravitational potential on scales larger than the horizon size at decoupling. Since many quantities, most notably the density contrast ($\delta\rho/\rho$), depend on the choice of gauge under these conditions (Bardeen 1980; Kodama & Sasaki 1984; Mukhanov, Feldman, & Brandenberger 1992), we must be careful to insure that our results are gauge-invariant. We address the gauge issue here.

The Newtonian gravitational potential has a natural gauge-invariant generalization, usually denoted by an upper case $\Phi$ [called $\Phi_H$ by Bardeen (1980)] to distinguish it from the usual Newtonian potential ($\phi$). The gauge-invariant Sachs-Wolfe formula linearly relates temperature fluctuations to $\Phi$, so our results hold true if $\Phi$ has the same statistics as $\phi$. We will show this to be true. However, a possible source of confusion arises. The gauge invariant generalization of the density perturbation is not unique. In fact the gauge-invariant density fluctuation variable (called $\epsilon_g$ by Bardeen 1980) which is used by Mukhanov *et al.* (1992), reduces (in flat space) to $-2\Phi$ on scales larger than the horizon during the matter dominated era, which seems to imply that there is no difference between the density and potential field, hence apparently nullifying our conclusions.

One way to see that our conclusions are correct is to use a different generalization of the density fluctuation called $\epsilon_m$. This is the density fluctuation relative to the space-like hypersurface which represents the local matter rest frame everywhere. This is the most natural variable for looking at matter perturbations. With this variable the usual Newtonian-like relation

$$\nabla^2 \Phi = 4\pi G \bar{\rho} \epsilon_m \qquad (A1)$$

holds for all scales, including suprahorizon scales. Thus no problem arises as the relation between $\epsilon_m$ and $\Phi$ does not change with scale. The apparent contradiction stems from the fact that when one constructs the variable $\epsilon_g$, one cancels out the gauge dependence by adding something which contains the potential itself. On subhorizon scales all gauge-invariant definitions of the density contrast converge to the same answer. On suprahorizon scales, however, the relationship between the two definitions mentioned here is

$$\epsilon_g - \epsilon_m = -2\left[1 - \frac{K}{\Omega a^2 H^2}\right]\left[\frac{1}{H}\frac{d\Phi}{dt} + \left(1 - \frac{K}{a^2 H^2}\right)\Phi\right], \qquad (A2)$$

where $K$ is the scalar curvature, $a$ is the scale factor, $\Omega$ is the fraction of the critical density, $H$ is the Hubble parameter, and $t$ is the time since the big bang. In flat space, ($K = 0$), the potential $\Phi$ for perfect fluids is constant, regardless of whether the universe is dominated by matter or radiation. Hence eq. (A2) reduces to

$$\epsilon_g - \epsilon_m \simeq -2\Phi. \qquad (A3)$$

Thus the reason the variable $\epsilon_g$ is $\sim -2\Phi$ on suprahorizon scales is that the $-2\Phi$ has been explicitly added on during the process of trying to cancel out the gauge dependence of $\delta\rho/\rho$.

With this distinction between different definitions clear, we now argue that the choice of variable has nothing to do with the argument put forth in the main part of the paper. On subhorizon scales, all gauge-invariant generalizations will reduce to the usual Newtonian $\delta\rho/\rho$. Thus, if the density fluctuations responsible for the present subhorizon structures are indeed non-Gaussian in character, the present potential field $\Phi$ must have the near Gaussian distribution under the conditions specified in this paper. If we extrapolate $\Phi$ back to the time of recombination, $\Phi$ will still exhibit this Gaussian behavior, since the potential is approximately constant in time. It is then irrelevant which density fluctuation variable we use to describe the density perturbations at recombination. The potential $\Phi$, and hence the temperature fluctuations, will have the Gaussian distribution under the conditions outlined in the text.

# REFERENCES

Bardeen, J.M. 1980, Phys Rev D, 22, 1882

Bardeen, J.M., Bond, J.R., Kaiser, N., & Szalay, A.S. 1986, ApJ, 304, 15

Bennett, D.P. & Rhie, S.H. 1993, ApJ, 406, L7

Coulson, D., Ferreira, P., Graham, P., & Turok, N. 1994, Nature, 368, 27

Feller, W. 1971, *An Introduction to Probability Theory and Its Applications*, Vol. II, (New York: Wiley)

Gorski, K.M., et al. 1994, ApJ, submitted (COBE Preprint 94-08)

Hinshaw, G., et al. 1994, ApJ, submitted (COBE Preprint 93-12)

Ibragimov, I.A., & Linnik, Yu. V. 1971, *Independent and Stationary Sequences of Random Variables*, (Groningen: Wolters-Noordhoff)

Kodama, H., & Sasaki, M. 1984, Prog Theor Phys, 74, 1

Kung, J.H. 1993, Phys Rev D, 47, 409

Lucchin, F. & Matarrese, S. 1988, ApJ, 330, 535

Luo, X. 1994a, Phys Rev D, 49, 3810

Luo, X. 1994b, ApJ, 427, L71

Luo, X., & Schramm, D.N. 1993, Phys Rev Lett, 71, 1124

Moscardini, L., Matarrese, S., Lucchin, F., & Messina, A. 1991, MNRAS, 248, 424

Moessner, R., Perivolaropoulos, L., & Brandenberger, R. 1994, ApJ, in press

Mukhanov, V.F., Feldman, H.A., and Brandenberger, R.H. 1992, Phys Rep, 215, 203

Peebles, P.J.E. 1980, *The Large-Scale Structure of the Universe* (Princeton: Princeton University Press)

Peebles, P.J.E. 1983, ApJ, 274, 1

Pen, U.-L., Spergel, D.N., & Turok, N. 1994, Phys Rev D, 49, 692

Scargle, J.D. 1981, ApJ Suppl, 45, 1

Scherrer, R.J. 1992, ApJ, 390, 330

Scherrer, R.J., & Bertschinger, E. 1991, ApJ, 381, 349

Scherrer, R.J., Melott, A.L., & Shandarin, S.F. 1991, ApJ, 377, 29

Smoot, G.F., et al. 1992, ApJ, 396, L1

Smoot, G.F., et al. 1994, ApJ, submitted (COBE Preprint 94-03)

Srednicki, M. 1993, ApJ, 416, L1

Wright, E.L., et al. 1992, ApJ, 396, L13

Wright, E.L, Smoot, G.F., Bennett, C.L., & Lubin, P.M. 1994, ApJ, submitted (COBE Preprint 94-02)

**FIGURE CAPTIONS**

Fig. 1: (a) The distribution of densities in the model given by the Wold representation: $\delta(\mathbf{r}) = \int f(|\mathbf{r} - \mathbf{r}'|)\Delta(\mathbf{r}')d^3\mathbf{r}'$, where $\Delta(\mathbf{r})$ is an uncorrelated field with a gamma distribution, and $f$ is chosen to give a Zel'dovich power spectrum $[P(k) \propto k]$ for the density field. Solid curve is a Gaussian distribution with the same mean and variance. (b) The distribution of potentials for the same model. Solid curve is a Gaussian distribution with the same mean and variance.

Fig. 2: As Fig. 1, for the power spectrum $P(k) \propto k^2$: (a) Distribution of densities. (b) Distribution of potentials.